\documentclass[conference]{IEEEtran}
\IEEEoverridecommandlockouts
\usepackage{amsmath,amssymb,amsfonts}
\usepackage{algorithmic}
\usepackage{graphicx}
\usepackage{textcomp}
\usepackage{xcolor}
\usepackage[style=numeric-comp, sorting=none]{biblatex}
\def\BibTeX{{\rm B\kern-.05em{\sc i\kern-.025em b}\kern-.08em
    T\kern-.1667em\lower.7ex\hbox{E}\kern-.125emX}}
\begin{document}

\title{Same Feedback, Different Source: How AI vs. Human Feedback Attribution and Credibility Shape Learner Behavior in Computing Education
}


\author{\IEEEauthorblockN{Caitlin Morris}
\IEEEauthorblockA{\textit{MIT Media Lab} \\
Cambridge, MA, USA \\
camorris@media.mit.edu}
\and
\IEEEauthorblockN{Pattie Maes}
\IEEEauthorblockA{\textit{MIT Media Lab} \\
Cambridge, MA, USA \\
pattie@media.mit.edu}
}

\thanks{This work has been submitted to the IEEE for possible publication. 
Copyright may be transferred without notice, after which this version may no longer be accessible.}

\maketitle

\begin{abstract}
As AI systems increasingly take on instructional roles — providing feedback, guiding practice, evaluating work — a fundamental question emerges: does it matter to learners who they believe is on the other side? We investigated this using a three-condition experiment (N=148) in which participants completed a creative coding tutorial and received feedback generated by the same large language model, attributed to either an AI system (with instant or delayed delivery) or a human teaching assistant (with matched delayed delivery). This three-condition design separates the effect of source attribution from the confound of delivery timing, which prior studies have not controlled. Source attribution and timing had distinct effects on different outcomes: participants who believed the human attribution spent more time on task than those receiving equivalently timed AI-attributed feedback (d=0.61, p=.013, uncorrected), while the delivery delay independently increased output complexity without affecting time measures. An exploratory analysis revealed that 46\% of participants in the human-attributed condition did not believe the attribution, and these participants showed worse outcomes than those receiving transparent AI feedback (code complexity d=0.77, p=.003; time on task d=0.70, p=.007). These findings suggest that believed human presence may carry motivational value, but that this value depends on credibility. For computing educators, transparent AI attribution may be the lower-risk default in contexts where human attribution would not be credible.
\end{abstract}


\section{Introduction}
As AI systems take on roles traditionally held by human instructors, such as providing feedback and evaluating work, a fundamental question arises: does it matter to learners who they believe is on the other side? When a student receives feedback on their code, does believing it came from a human change how they work, what they produce, or how they experience the task?

These questions have practical urgency. As computing educators integrate AI feedback tools alongside human instruction — and increasingly use large language models (LLMs) to help draft or refine their own feedback — they need to know whether learners respond differently to feedback they believe came from a person versus an AI, independent of the feedback itself. Prior research on source attribution effects is mixed: a recent meta-analysis found no significant overall difference in learning outcomes between AI and human feedback \cite{Kaliisa2025-rs}, but studies examining perception find that identical content is evaluated differently depending on its attributed source \cite{Lipnevich2008-eu, Rubin2025-lr}. Critically, existing studies typically confound source attribution with other delivery characteristics, particularly timing. In naturalistic comparisons, AI feedback arrives instantly while human feedback involves a delay, making it unclear whether observed differences reflect attribution or the experience of waiting. Additionally, no prior work has examined how the \emph{credibility} of feedback attribution influences its effects — an increasingly relevant question not only as learners become more AI-literate, but as instructors themselves begin using LLMs to write or augment feedback that they deliver under their own name. In both cases, learners may suspect that ostensibly human feedback was actually generated by AI, and what happens when that suspicion arises is unknown.

We designed a three-condition experiment to address both gaps. Participants completed a self-paced creative coding tutorial using p5.js \cite{Processing-Foundation2014-hv}, receiving feedback at four checkpoints. All feedback was generated by Claude Sonnet 4 using an identical prompting template that specified tone, structure, and a consistent balance of encouragement and specific correction — tailored to each participant's specific code submission but otherwise equivalent across conditions. Participants were randomly assigned to one of three conditions:

\begin{itemize}
\item AI-Instant: Feedback delivered after 5 seconds, labeled as AI-generated
\item AI-Delayed: Feedback delivered after 60 seconds with a processing animation, labeled as AI-generated
\item TA-Delayed: Feedback delivered after 60 seconds with a review animation, labeled as coming from a human teaching assistant (TA)
\end{itemize}

Comparing AI-Delayed to TA-Delayed isolates source attribution with timing held constant. Comparing AI-Instant to AI-Delayed isolates the effect of delivery timing with attribution held constant. The study was pre-registered \cite{Morris2026-pn} following a pilot study (N=25, also pre-registered \cite{Morris2026-hx}) that identified both the behavioral effect of interest and the potential timing confound. We address three research questions: 
\begin{itemize}
\item RQ1: When the feedback generation and structure is held constant, does source attribution (AI vs. human) affect learner behavior — and can this effect be separated from the confound of delivery timing?
\item RQ2: Does source attribution interact with feedback type, shaping responses to technical versus creative feedback differently?
\item RQ3 (exploratory): Does the credibility of human attribution moderate its behavioral effects, and do non-credible attributions produce outcomes worse than transparent AI?
\end{itemize}

The primary confirmatory contribution is the effect of believed source attribution on learner behavior (RQ1), with the timing contrast serving as a methodological control. RQ2 is a secondary confirmatory test and RQ3 is exploratory. Our findings address both the confirmatory and exploratory questions. Source attribution and delivery timing have separable effects on different aspects of learner behavior: believed human attribution increased time on task and active focus (\emph{process} measures), while the delivery delay independently increased code length and complexity (\emph{output} measures). These are distinct and non-overlapping effects. An exploratory analysis of attribution credibility suggested a further pattern: nearly half of participants in the human-attributed condition did not believe the attribution, and these participants showed the worst outcomes of any group — producing less complex code and shorter programs than those who received transparently attributed AI feedback, as well as reporting less interest. This exploratory finding, if confirmed, has practical implications for how source claims are framed in hybrid learning environments.

\section{Theoretical Foundations}

Our study draws on three complementary theoretical perspectives that motivate the study design and offer predictions about why source attribution could affect learner behavior — and what might happen when it is not believed.

\subsection{Social Presence Theory}

Social presence theory suggests that communication media vary in the degree to which they convey the sense that another person is present in the interaction \cite{Parker1978-ss}. In educational contexts, higher social presence is associated with greater satisfaction and perceived learning \cite{Richardson2019-to, Richardson2017-st}. The most intuitive explanation for why human-attributed feedback might increase learner effort is through social presence: believing a human instructor is reviewing your work may create a sense of being observed, evaluated, and accountable, activating social motivations that an AI label does not elicit. This account predicts that behavioral differences between conditions should be accompanied by differential self-reported social presence — participants who believe a human is reviewing their work should feel more watched and more motivated to impress.

\subsection{Self-Determination Theory and Intrinsic Motivation}

Self-determination theory (SDT) \cite{Ryan2000-hy} identifies three basic psychological needs — autonomy, competence, and relatedness — that support intrinsic motivation. In educational settings, satisfaction of these needs promotes deeper investment and more autonomous forms of motivation \cite{Niemiec2009-gy}. SDT offers an alternative account: believing that a real human has engaged with your work may support the need for relatedness — the sense that one's efforts are seen and valued by another person — fostering intrinsic motivation and sustained voluntary investment. Crucially, this account does not require surveillance or evaluation pressure; the mechanism is the perceived authenticity and relational quality of the interaction, not the sense of being watched.

This distinction becomes critical when attribution is not believed. If social presence drives the effect, non-credible attribution should simply reduce the sense of being watched, returning behavior to the AI baseline. If relational authenticity drives the effect, non-credible attribution may undermine intrinsic motivation more severely — producing outcomes worse than transparent AI, because the learner now perceives a disingenuous interaction rather than simply a non-human one.

\subsection{Trust, Authenticity, and Credibility}

Research on trust in human-computer interaction suggests that users form expectations about system behavior based on how the system is presented, and that violations of these expectations can produce disproportionate negative reactions \cite{Madhavan2007-pf}. The trust violation literature documents "betrayal aversion" — the finding that negative outcomes attributed to intentional betrayal are experienced more negatively than equivalent outcomes attributed to chance or incompetence \cite{Bohnet2004-pv}. When users discover that a system misrepresented its nature, they may not simply correct their expectations; they may withdraw investment as a form of reactance \cite{Brehm1981-qo}. Applied to our context, the trust violation framework predicts the same asymmetric pattern as the SDT account: non-credible human attribution should produce worse outcomes than transparent AI, not merely equivalent ones.

\section{Related Work}

\subsection{AI Feedback in Computing Education}
AI-powered feedback tools are rapidly entering computing education. LLMs have been used to generate formative programming feedback \cite{Kiesler2023-jq}, provide automated code review and grading \cite{Phung2023-cq}, and serve as conversational tutoring agents in introductory CS courses \cite{Kasneci2023-cs}. This growth is driven by the promise of scalability: personalized, immediate feedback on student code is difficult for instructors to sustain across large courses, and LLM-generated feedback can approximate it \cite{Kasneci2023-cs}. A recent meta-analysis by Kaliisa et al. synthesized 41 studies comparing AI and human feedback across educational domains \cite{Kaliisa2025-rs}, finding no significant overall difference in learning performance but high variability and a trend toward students perceiving human feedback more favorably. The studies in this analysis confound source with feedback content, making it difficult to isolate attribution effects. Nazaretsky et al. began to address this by showing that students rated the same feedback less favorably after learning it was AI-generated \cite{Nazaretsky2024-iw}, demonstrating a source credibility effect — though their design did not test naturally occurring belief, and measured perception rather than behavioral outcomes.

This confound is especially relevant in computing education, where AI feedback tools are often deployed alongside human instructors and TAs. Students may receive AI-generated code feedback in one context and human feedback in another — and may form expectations about the relative value of each. Computing students are also likely to be among the most AI-literate populations \cite{Long2020-xo}, with growing familiarity with LLM-generated text \cite{Clark2021-jb} that may increase skepticism about claimed human authorship in online contexts. Understanding whether the attributed source of feedback independently shapes learner behavior, separate from feedback content and delivery timing, is therefore a practical design question for hybrid computing courses.

\subsection{Perception, Behavior, and the Feedback Gap}
Research on feedback effectiveness has long recognized a gap between how learners evaluate feedback and what they do with it \cite{Hattie2007-zb, Evans2013-bv}. Hattie and Timperley established that feedback is among the most powerful influences on learning, but its effectiveness depends on how learners receive and act on it — not just on its quality or their ratings of it \cite{Hattie2007-zb}. Research on the "feedback gap" has found that the relationship between perceived helpfulness and actual impact of feedback is weak, suggesting that self-report measures of feedback quality may not capture the motivational aspects of the feedback experience \cite{Evans2013-bv}.

Source attribution may be one factor that widens or narrows this gap. Lipnevich and Smith found that college students improved on essay tasks when given detailed feedback regardless of whether it was attributed to an instructor or a computer, with specificity of feedback carrying more weight on outcomes than grades or praise \cite{Lipnevich2008-eu}. However, Rubin et al. showed that identical empathic responses were rated as less empathic when labeled AI-generated, indicating that source labels shape qualitative perception \cite{Rubin2025-lr}. Together, these studies suggest that attribution may affect perceptual evaluation and behavioral investment through different pathways; our study is designed to test this dissociation.

Following broader research that engagement is a multidimensional construct \cite{Fredricks2004-po}, we measure behavioral process indicators (time on task, iterative testing), output measures (code length, code complexity), and perception measures (helpfulness ratings) separately. This allows us to test whether source attribution affects what learners \emph{do} and what they \emph{produce}, not just what they \emph{report}.

\section{Method}

\subsection{Participants and Design}
We recruited 150 participants through Prolific, using a two-stage process. A screening survey assessed programming experience, interest in learning creative coding, and baseline attitudes toward AI tools and human teaching assistants. These attitude measures were collected at screening — before condition assignment — to prevent contamination from the experimental manipulation. Participants who met eligibility criteria (interest in learning programming, English fluency, desktop/laptop access, non-advanced experience) were invited to the main study. Compensation was \$16 for approximately 60 minutes. Two participants were excluded from analysis for non-participation (submitting no code), yielding an analytic sample of N=148 (50.3\% female; mean age 37.3, SD=11.6). The sample was mixed between novice and introductory-level experience: 78 participants had no prior coding experience but had an interest in learning, and 70 had some introductory experience.

\begin{table*}[htbp]
\caption{Feedback Samples}
\begin{center}
{\renewcommand{\arraystretch}{1.5}%
\begin{tabular}{|c|c|p{0.32\linewidth}|c|p{0.32\linewidth}|c|}
\hline
\textbf{Group} & \textbf{\textit{Module}}& \textbf{\textit{Technical Feedback}}& \textbf{\textit{ Rating}} & \textbf{\textit{Creative Feedback}} & \textbf{\textit{ Rating}} \\
\hline
AI-Delayed& 
Color& 
You have two 'function draw()' declarations, which means only the second one will run. Try combining your code into a single draw() function 
 &  4 & 
You've created a charming house scene with good use of basic shapes - the triangle roof, rectangular house, and circular sun work well together. Consider exploring different color combinations for a more natural palette.
 &  4 \\
\hline
AI-Delayed& 
Interaction& 
There are some syntax errors in your keyPressed() function - keyCode should be compared to actual key codes (numbers) rather than letters, and the variables x and y aren't defined in your code
 &  5 &
What if pressing different keys could change the time of day by shifting the background color, or clicking the mouse could turn the house lights on and off by changing the window colors?
 & 3
\\
\hline
TA-Delayed& 
Color& 
Your triangle has all six coordinates set to (200,200) - this creates a single point rather than a triangle. Try using three different coordinate pairs to form the triangle's corners.
 &  4 &
You've got a good start with three ellipses positioned in different areas - what if those became part of something larger, like a face with two eyes, or three planets in a solar system? & 3
\\
\hline
TA-Delayed& 
Interaction& 
Your mousePressed() function correctly changes the size variable when clicked, but since that variable isn't used to draw anything, you won't see the effect - try using it in your ellipse size. &  5 &
Consider experimenting with different colors for variety - right now you're using pure primary colors, but what if you tried mixing different RGB values to create more natural or interesting color combinations? & 3
\\
\hline
\end{tabular} }
\label{tab1}
\end{center}
\end{table*}

The study used a 3 (Condition: AI-Instant, AI-Delayed, TA-Delayed; between-subjects) x 2 (Feedback Type: technical, creative; within-subjects) mixed factorial design. After exclusions, condition sizes were: AI-Instant (n=49), AI-Delayed (n=49), TA-Delayed (n=50).

\begin{figure}[tbp]
\centering
\includegraphics[width=0.4\textwidth]{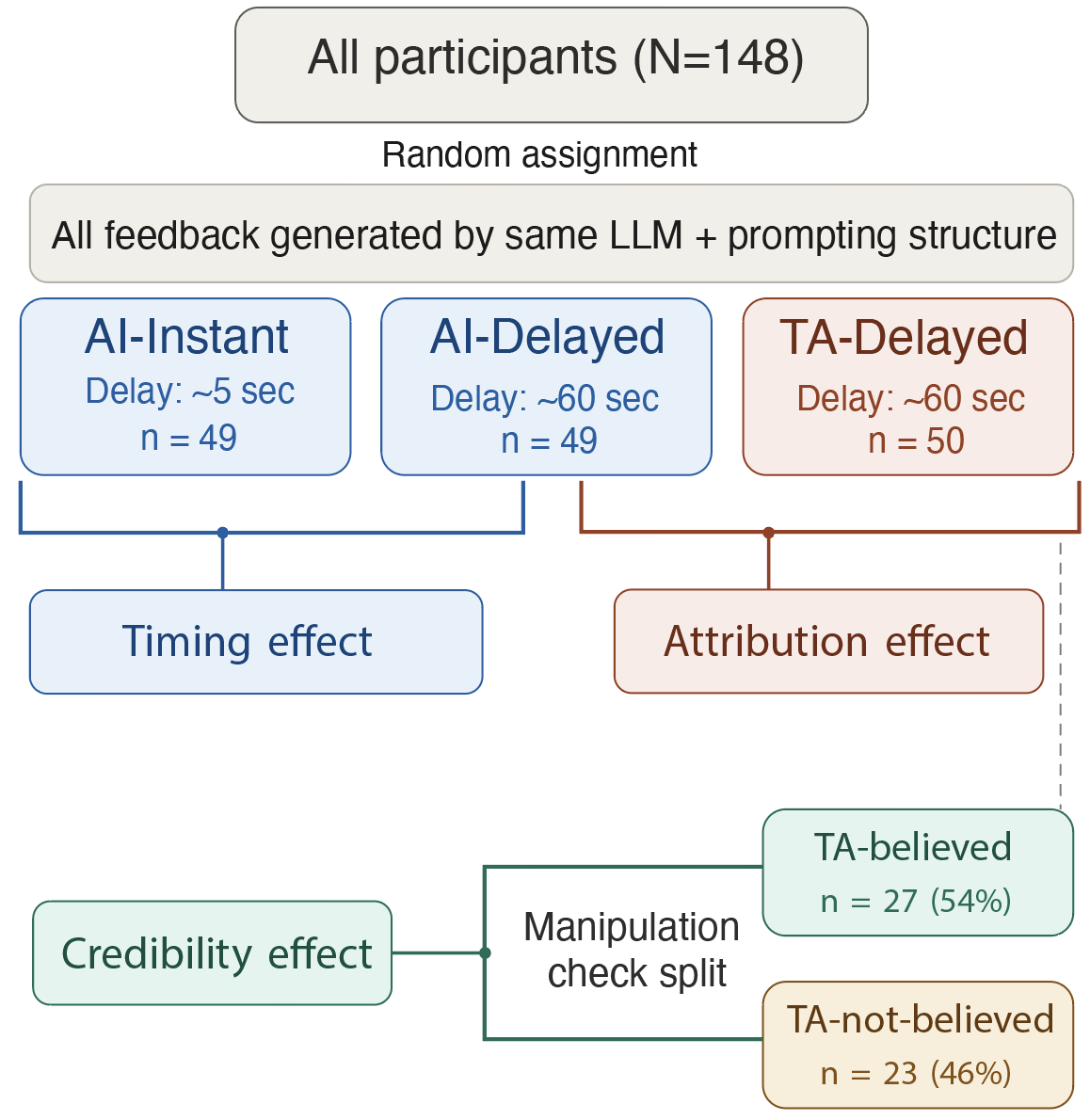}
\caption{Experiment design}
\label{fig}
\end{figure}

\subsection{Materials and Procedure}

Participants completed a self-paced creative coding tutorial using p5.js, a JavaScript library for creating visual and interactive graphics through code, with four progressive modules (shapes, colors, interaction, mini-project). At each checkpoint, participants submitted their code and received both technical feedback (on code structure and practices) and creative feedback (on visual design and exploration). All feedback was generated by Claude Sonnet 4 using the same prompting structure, tailored to each participant's specific submission. Because each participant wrote different code, the specific feedback content varied across individuals; what was held constant was the generation process — the same model, prompting template (equivalent mix of encouragement and specific notes for improvement), tone, and format across all conditions. Table \ref{tab1} shows example feedback from two participants, representing AI-Delayed and TA groups, to illustrate the personalization, specificity, and consistency of tone across submissions.  Only the attributed source and delivery timing differed between conditions.

After each checkpoint, participants rated the helpfulness of technical and creative feedback separately on 5-point Likert scales (ratings of example feedback are shown in Table \ref{tab1}). After completing all four modules, participants completed a post-tutorial survey covering feedback perception, social presence and accountability, motivation factors, feedback influence, learning and enjoyment, and a manipulation check.

The study included an IRB-approved deception component: all participants in the TA-Delayed condition were told their feedback came from a human teaching assistant, when in fact it was generated by the same LLM as in the AI conditions. Participants were debriefed after the manipulation check.

\subsection{Measures}

Analysis measures include the following:

\begin{itemize}
    \item Process indicators (logged automatically): Time on task per module (seconds), time with the tutorial browser window in focus, and code execution frequency (number of times the participant ran their code, indicating iterative testing behavior).
    \item Output measures: Code length and a composite code complexity score (0-8). The complexity score was computed automatically from each code submission by a scoring script that assessed eight binary or normalized dimensions: \emph{shape variety} (number of distinct shape types used), \emph{shape volume} (total shapes drawn), \emph{color usage} (number of unique fill/stroke colors), \emph{style function usage} (e.g. strokeWeight, noStroke), \emph{interaction features} (e.g. mouseX, keyPressed), \emph{variable declarations}, \emph{control flow structures} (loops, conditionals), and \emph{advanced programming features} (e.g., transforms, trigonometry, custom functions). Each dimension was capped at a maximum contribution of 1.0 and the eight dimensions were summed to produce a score ranging from 0 to 8. Because prior programming experience predicted raw complexity scores, we adjusted scores for experience by removing the linear effect of experience level and centering on the sample mean, so that comparisons reflect differences in effort rather than baseline skill.
    \item Perception measures: Per-checkpoint helpfulness ratings (5-point Likert for technical and creative feedback separately) and post-survey items on feedback perception (genuineness, understanding, perceived effort), social presence and accountability (feeling watched, judged, impression motivation, appreciation), motivation factors (intrinsic interest, evaluation-driven effort, autonomous effort), feedback influence, and learning/enjoyment.
    \item Manipulation check: Survey item asking whether the participant believed the feedback came from the attributed source, AI or human TA (5-point Likert). Participants in the TA condition scoring $ \leq $ 2 were classified as non-believers.
\end{itemize}

\subsection{Analytic Approach}

Behavioral measures were collected at each of four checkpoints, yielding repeated observations per participant. For the Source × Feedback Type interaction (RQ2), we fit linear mixed-effects models (LMMs) with condition, feedback type (effects-coded), and their interaction as fixed effects and participant as a random intercept, using the statsmodels MixedLM implementation. For behavioral indicators (RQ1), we fit LMMs with condition and checkpoint (linear) as fixed effects and participant as a random intercept for the repeated-measures structure; planned contrasts were then computed on participant-level means using independent-samples t-tests with Cohen's d for effect sizes. The three planned contrasts (Attribution, Timing, Ecological) were pre-registered and not corrected against each other because they test distinct, pre-specified hypotheses. Within each contrast, we applied a pre-registered Bonferroni correction across the five behavioral DVs ($\alpha$= .05/5 = .01) and report both corrected and uncorrected p-values. This correction is conservative given that several DVs are correlated (e.g., time on task and time in focus), making the effective number of independent tests less than five.

Primary analyses used the pre-registered three-condition framework (excluding TA non-believers, N=125). Because the TA condition is filtered to only those who believed or were ambiguous about the human attribution, the confirmatory attribution contrast estimates the effect of \emph{believed} human attribution, not just condition assignment. The exploratory 4-group analysis included all 148 participants, splitting the TA condition by believability. We verified that TA-believed (n=27) and TA-not-believed (n=23) groups did not differ on any baseline measure. 

Post-survey items were analyzed with one-way ANOVAs across the four groups and pairwise t-tests for the TA-believed vs. TA-not-believed comparison. These mechanism analyses are exploratory.

\section{Results}

\subsection{Manipulation Check}
Of 50 TA-condition participants, 27 (54\%) reported believing or being ambiguous about the human attribution (manipulation check $ \geq $ 3), while 23 (46\%) reported disbelief. Responses were strongly bimodal: 24 participants rated 4-5 (clear belief), while 23 scored 1-2 (clear disbelief), with only 3 participants at the ambiguous midpoint of 3. Because so few participants fall at the threshold, excluding or reassigning the 3 midpoint participants does not meaningfully change the pattern of results. The two subgroups did not differ significantly on any pre-study measure: coding experience (d=-0.03, p=.92), AI tool usage (d=0.25, p=.15), opinion of AI (d=0.18, p=.60), opinion of human TAs (d=0.14, p=.60), age (d=-0.002, p=.89), or sex (X2=1.29, p=.26). This baseline equivalence strengthens the interpretation that behavioral differences between subgroups reflect the credibility of the attribution rather than pre-existing individual differences. Both AI conditions showed high source believability (M=4.45 on the 5-point manipulation check).

The non-credibility rate likely reflects the high AI literacy of the Prolific population and may have been amplified by pre-survey AI attitude items collected 1–2 days prior (see Limitations for further discussion).

\subsection{Source Attribution and Feedback Type (RQ2)} Using the AI-Delayed vs. TA-believed contrast with timing controlled, the interaction between source attribution and feedback type on helpfulness ratings was non-significant (p=.22). Technical feedback was rated higher than creative feedback across all conditions, but this gap did not differ by source. This null result replicates the pilot finding and provides robust evidence against the hypothesis that learners trust AI differently for technical vs. creative feedback.

\subsection{Attribution and Timing Effects on Behavior (RQ1)} 
\subsubsection{Attribution (AI-Delayed vs. TA-believed, timing controlled)} Participants who believed the human attribution showed greater time on task (M=948s vs. 741s; d=0.61, p=.013) and time in focus (M=850s vs. 662s; d=0.56, p=.023). Both effects are significant at the uncorrected $\alpha$=.05 threshold but not at the pre-registered Bonferroni correction ($\alpha$=.01); we report both and interpret the pattern together with effect sizes. Code execution frequency, code length, and code complexity were in the predicted direction but did not reach significance at this sample size (code runs: d=0.29, p=.235; code length: d=0.33, p=.171; code complexity: d=0.22, p=.357). The attribution effect thus manifests primarily in process indicators — how long and how actively participants worked — rather than in output measures at this contrast. As we report below, output measures differentiate strongly in the believability analysis.

\subsubsection{Timing (AI-Instant vs. AI-Delayed, attribution controlled)} There were no significant differences on process indicators: time on task (d=0.18, p=.365), time in focus (d=0.10, p=.610), or code execution frequency (d=0.01, p=.950). The delivery delay did not cause participants to spend more time or test more iteratively, confirming that the attribution effects on process are attributable to source, not timing. However, the delay did independently affect output measures: AI-Delayed participants produced more complex programs (d=0.55, p=.008, significant at corrected $\alpha$=.01) and wrote longer code (d=0.43, p=.038, significant at uncorrected $\alpha$=.05 but not corrected $\alpha$=.01). Source attribution and delivery timing thus suggest distinct effects — attribution shapes the work process while timing shapes what participants produce, possibly by creating a natural pause for reflection before continuing.

\begin{figure*}[tbp]
\centerline{\includegraphics[width=1.0\textwidth]{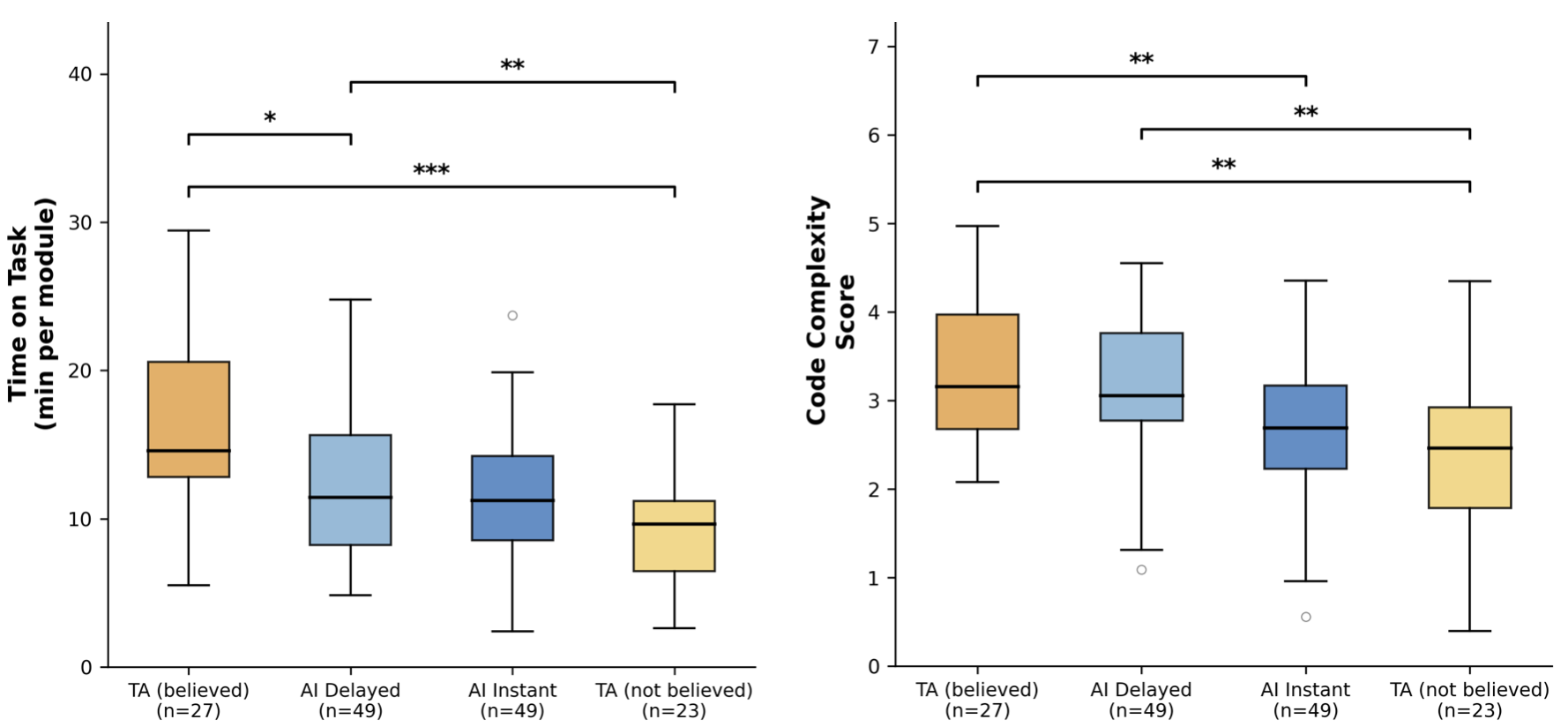}}
\caption{Behavioral measures by group. (L) Time on task in minutes per module. (R) Code complexity score, experience-adjusted.}
\label{boxplot-behavioral}
\end{figure*}

\subsubsection{Helpfulness ratings} There were no significant differences in overall helpfulness ratings across conditions (AI: M=4.09, AI-Delayed: M=4.21, TA-believed: M=4.44; TA vs. AI-Delayed p=.155). The behavioral effect of source attribution does not manifest in explicit evaluations of feedback quality, consistent with feedback gap literature \cite{Evans2013-bv}, in which perceived helpfulness and actual behavioral response to feedback diverge.

For RQ1, process measures show consistent, medium-sized differences in the predicted direction that reached uncorrected significance: believed human attribution increases time on task and active focus, independent of timing (d=0.56–0.61).

\subsection{The Role of Credibility (RQ3)}
Our pre-registered plan specified that participants failing the manipulation check would be flagged and analyses reported with and without them. The magnitude of the believability split and the behavioral divergence between subgroups prompted the more detailed exploratory analysis reported here.

When all 148 participants are included with the TA condition split by believability, a clear hierarchy emerges across behavioral indicators (Fig. \ref{boxplot-behavioral}): \textbf{TA-believed $>$ AI conditions $>$ TA-not-believed.} The former gap (TA-believed $>$ AI conditions) reflects the benefit of believed human presence. TA-believed participants outperformed both AI conditions on time on task (d=0.61–0.81 vs. AI groups), code length (d=0.67), and code complexity (d=0.75).

The latter gap (AI conditions $>$ TA-not-believed) is more substantial. TA-not-believed participants did not simply revert to AI-condition levels; they fell below them. Compared to AI-Delayed participants (who received equivalently timed AI-attributed feedback), TA-not-believed participants produced significantly shorter programs (d=0.82, p=.002), less complex code (d=0.77, p=.003), and spent less time in focus (d=0.68, p=.009).

The full gap between TA-believed and TA-not-believed was very large across both process and output measures: time on task (d=1.28, p$<$.001), time in focus (d=1.17, p$<$.001), code complexity (d=0.96, p=.001), code length (d=0.94, p=.002), and code executions (d=0.67, p=.022) — despite these groups being indistinguishable on all baseline measures, randomly assigned to the same condition, and receiving feedback generated by the same process. 

Self-reported feedback behavior from the post-survey reinforced the behavioral data. TA-not-believed participants reported skimming feedback at nearly three times the rate of other groups (26\% vs. 7-10\%), reading carefully at the lowest rate (61\% vs. 74-84\%), and feeling motivated by feedback at the lowest rate (22\% vs. 31-52\%). They withdrew not just from the task but from the overall feedback loop. 

Open-ended comments reinforced this pattern. TA-believed participants more frequently expressed enthusiasm and gratitude (``One of the most enjoyable studies I have done on Prolific"; ``That was fun, I'm going to come back and learn more"), while TA-not-believed participants were more muted or critical (``It was okay"; "The feedback was pretty clearly from an AI, so I'm not sure if I was fooled"). Among AI-Delayed participants, two independently reported that the 60-second delay felt like a technical error (``At first I thought there was a bug"), reflecting expectations about immediacy of AI feedback.

This pattern is consistent with the SDT and trust violation predictions from our theoretical framework: non-credible attribution produced outcomes worse than transparent AI, more consistent with the authenticity account than with a simple social presence reduction model.

\subsection{Mechanism: Social Presence vs. Intrinsic Motivation}
Our theoretical framework offers different predictions about the mechanism underlying attribution effects. Our post-survey data allow a preliminary test.

\begin{figure}[tbp]
\includegraphics[width=0.5\textwidth]{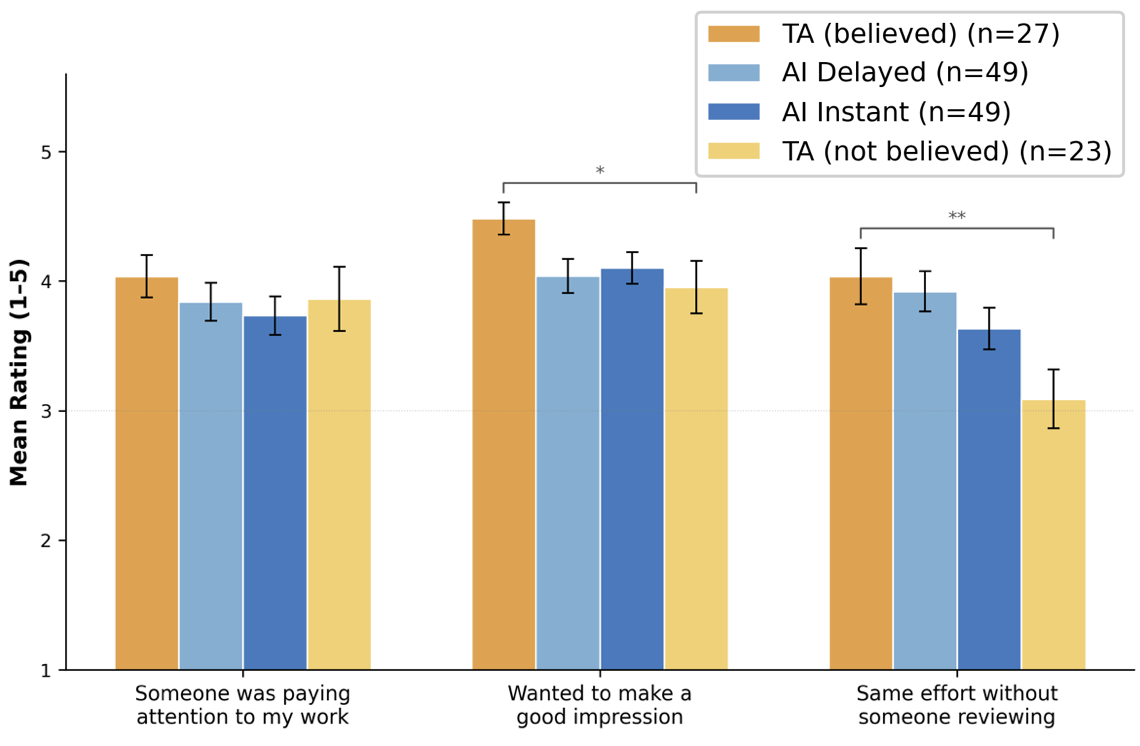}
\caption{Selected motivation survey items by condition.}
\label{motivation-factors}
\end{figure}

Social presence items were flat across all four groups. Feeling that someone was paying attention to their work (F=0.40, p=.75), feeling their work would be judged (F=0.56, p=.64), and wanting to show appreciation for the reviewer (F=0.52, p=.67) did not differ across conditions (Fig. \ref{motivation-factors}, left). Critically, TA-believed and TA-not-believed participants reported equivalent levels of social presence (d=0.13, d=0.06, and d=-0.06 respectively, all p$>$.60). External motivation — effort driven by feeling watched or evaluated — also showed no group differences (F=1.20, p=.31). All participants, regardless of attributed source or believability, reported feeling similarly observed and evaluated. This disagrees with the social presence account, which predicts that believed human attribution should increase feelings of being watched.

\begin{figure}[tbp]
\includegraphics[width=0.5\textwidth]{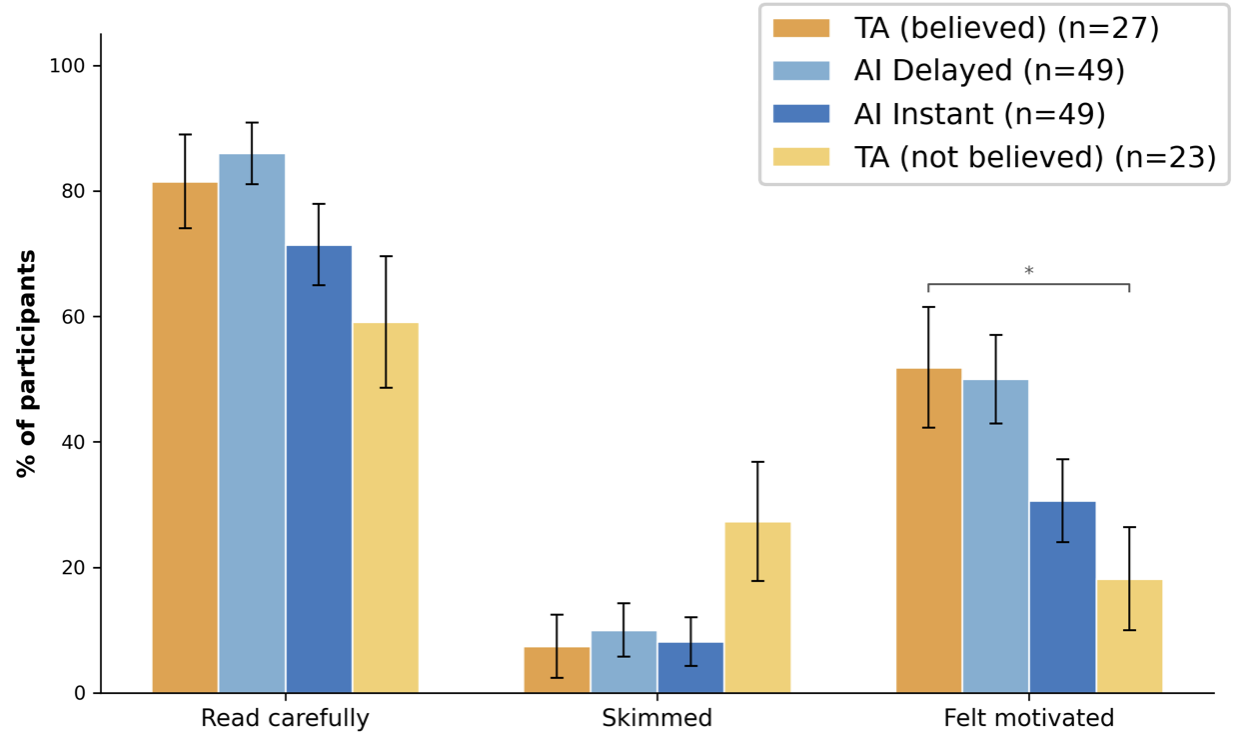}
\caption{Self-reported feedback behavior by condition.}
\label{feedback-interaction}
\end{figure}

Intrinsic motivation items show more differentiation. TA-believed participants reported the highest intrinsic interest in the creative coding tasks (M=4.52), while TA-not-believed participants reported the lowest (M=3.83) — lower than both AI conditions (M=4.22, 4.37). The difference between TA subgroups was large (d=0.80, p=.007). Autonomous effort (``I would have put in the same effort even without someone reviewing my code") was the strongest differentiator: TA-believed M=4.11 vs. TA-not-believed M=3.04 (d=1.01, p=.0008) (Fig. \ref{motivation-factors}, right). Believers' effort was autonomous and interest-driven; non-believers' effort was contingent and reduced. Impression motivation — wanting to make a good impression with their code — was the one social item that differentiated (d=0.67, p=.023), reflecting self-presentation motivation rather than surveillance (Fig. \ref{motivation-factors}, middle).

TA-believed participants also reported significantly higher enjoyment of the tutorial (M=4.81 vs. TA-not-believed: M=4.13; 4-group F=3.75, p=.012). This pattern is more consistent with the SDT/authenticity account than the social presence account. The behavioral hierarchy tracks intrinsic motivation and perceived authenticity rather than surveillance or evaluation pressure, though these initial findings require confirmation with validated multi-item scales.

These exploratory findings suggest an asymmetry for human-attributed feedback in this context. Among those for whom the human attribution was credible, outcomes were the best of any condition — a moderate-to-large advantage over transparent AI. Among those for whom it was not credible, outcomes were the worst of any condition — a moderate-to-large disadvantage relative to transparent AI. With a 46\% non-credibility rate in our sample, the average outcome of the human-attributed condition is pulled toward or below the AI-condition mean, despite the strong performance of the believed subgroup. If this pattern replicates, it implies that the benefit of human attribution is conditional on credibility; when credibility fails, the costs may outweigh the gains.

\section{Discussion}

\subsection{Attribution Effect on Process}

The attribution contrast suggests that perceived human involvement in feedback has motivational value. With timing controlled, participants who believed a human TA reviewed their work spent approximately 28\% more time on task and maintained more active focus than those receiving equivalently timed AI-attributed feedback. Post-survey data are consistent with the interpretation that this additional time reflects sustained investment: TA-believed participants reported the highest enjoyment of the tutorial and the highest intrinsic interest in the creative coding tasks. This does not appear to be a timing artifact: the timing contrast shows that the 60-second delay alone does not increase time on task or active focus.

This finding is consistent with SDT's relatedness account: believed human attribution may support the sense that one's work is seen and valued by another person, fostering intrinsic motivation. The TA-believed group's high scores on intrinsic interest and autonomous effort items are consistent with this interpretation, though the single-item nature of these measures warrants caution. The three-condition design reveals that source and timing affect different outcomes: attribution shapes the work process (time and focus), while the delivery delay shapes the work product (code length and complexity), possibly by creating a natural pause for reflection. Both are meaningful design levers, but they operate independently.

\subsection{Non-Credible Attribution and Behavioral Withdrawal}

The exploratory credibility analysis suggests that non-credible human attribution produces outcomes worse than transparent AI. Participants who suspected the TA attribution was false produced the least complex code and the shortest programs of any group — significantly less than participants who \emph{knew} they were receiving AI feedback. They also spent less time on task, less time in focus, and reported skimming feedback at nearly three times the rate of other groups. Because believability was not experimentally assigned, this finding should be interpreted as a suggestive pattern within a randomized design rather than a confirmed causal effect; however, the clean baseline equivalence between believer and non-believer subgroups strengthens the interpretation.

This pattern is consistent with both the SDT and trust violation frameworks outlined in Section II. From an SDT perspective, non-credible attribution may undermine autonomy (the learner perceives manipulation) and relatedness (the interaction feels disingenuous), deflating intrinsic motivation below what transparent AI would produce. From a trust violation perspective, perceived inauthenticity may trigger a withdrawal of investment that goes beyond what transparent AI produces — because transparent AI makes no relational claim to violate. We note that alternative causes for the differences may exist (see Limitations) and that confirmatory work manipulating credibility directly is needed.

\subsection{Social Presence vs. Intrinsic Motivation as Mechanism}
The null social presence results are inconsistent with the most common implicit explanation for source attribution effects in the literature. All groups felt similarly observed and evaluated, regardless of attributed source or believability. If replicated, this would be an important finding: the value of human presence in feedback may not operate through surveillance or evaluation pressure.

One explanation is that personalized feedback itself, regardless of source label, creates a baseline sense of being attended to. All participants received specific, tailored feedback on their work at every checkpoint. This feedback may have saturated social presence across conditions — consistent with Lipnevich and Smith's finding that detailed, specific feedback from both human and machine sources improved student performance \cite{Lipnevich2008-eu}. If so, social presence may matter for the difference between receiving feedback and receiving none, but not for the difference between AI-attributed and human-attributed feedback. However, these mechanism analyses are exploratory and based on single survey items, so they should be treated as suggestive directions rather than definitive conclusions.

\subsection{Design Implications for Computing Education}

These findings have practical relevance beyond the specific scenario of a system designed to impersonate a human. Instructors increasingly use LLMs to draft, augment, or refine feedback that they then deliver under their own name — a practice that improves efficiency but may risk the same credibility dynamic we observed. If students suspect that feedback labeled as coming from their instructor was actually generated by AI, the motivational benefit of perceived human involvement could diminish or reverse. This concern extends beyond education: growing public skepticism about whether human-attributed content is genuinely human-authored is a broader phenomenon that our findings may speak to.

Our confirmatory results suggest that where human involvement in feedback is real and verifiable — such as in established courses with known TAs — there is a behavioral benefit worth preserving. Our data also suggest that transparent AI attribution produces adequate behavioral outcomes and may be the lower-risk default in contexts where human attribution would not be credible. The exploratory credibility analysis raises the possibility that the risk lies specifically in the gap between claim and credibility, though this finding requires confirmatory replication before strong design recommendations can be drawn from it.

\subsection{Limitations}
The believability split is post-hoc, not randomly assigned; despite clean baseline equivalence on all pre-study measures, causal claims about believability should be interpreted cautiously. The TA subgroups (n=27 believed, n=23 not believed) are small, limiting precision. Alternative explanations for the non-believed group's underperformance — including dispositional skepticism, reactance to perceived deception, or aversion to waiting when AI is suspected — cannot be fully ruled out. The specificity of the behavioral pattern (non-believers skimmed feedback more and reported lower motivation, rather than performing uniformly worse) is more consistent with a feedback-interaction account, but direct experimental manipulation of credibility is needed.

The Prolific population likely has higher AI literacy than typical classroom learners, potentially inflating the non-credibility rate. Two pre-survey items about AI attitudes may have primed TA-condition participants to consider AI as a possible source, though the 1–2 day gap between screening and the main study mitigates this concern. The rate is context-dependent: in classroom settings with known instructors, credibility would presumably be higher and the human-presence benefit correspondingly more reliable.

The mechanism analysis (Fig. \ref{motivation-factors}) is exploratory and based on single survey items rather than validated multi-item scales. These findings should be treated as preliminary evidence suggesting a direction for further investigation, not as a definitive test of competing theoretical accounts.

The creative coding context, while well-suited to examining both technical and creative feedback, is a specific domain. Findings may differ in purely technical programming courses, in higher-stakes assessment contexts, or in domains with stronger prior expectations about AI competence.

\subsection{Future Work}
Future work should test these effects in classroom settings where instructor identity is established and verifiable, predicting higher credibility and a more robust human-presence benefit. Experimentally manipulating credibility — for example, by varying the quality of the human framing — would allow causal claims about believability rather than relying on post-hoc splits. Examining hybrid framings such as "AI-assisted feedback reviewed by your instructor" may capture some human-presence benefit while maintaining authenticity. The mechanism analysis should be extended with validated multi-item scales and mediation analysis.

\section{Conclusion}
Source attribution and delivery timing have separable effects on learner behavior in AI-augmented learning environments. Participants who believed their feedback came from a human instructor spent more time on task and stayed more focused than those receiving equivalently timed AI-attributed feedback (medium effect sizes, significant at the uncorrected threshold), while the delivery delay independently increased output complexity. This three-condition design provides the first evidence that these effects are distinct, addressing a timing confound present in prior work.

An exploratory analysis of attribution credibility suggested a further pattern: when the sense of human presence was not believed, outcomes were worse than transparent AI, not merely equivalent. Preliminary mechanism data are more consistent with intrinsic motivation and perceived authenticity than with social presence or evaluation pressure, though these exploratory findings require confirmation.

These findings are relevant not only for systems explicitly designed to simulate human feedback, but for the increasingly common practice of instructors using AI to draft feedback that they deliver as their own. As learners grow more familiar with AI-generated content, the assumption that human attribution will be believed cannot be taken for granted. For computing educators, genuine human involvement in feedback appears to carry real value — but where that involvement is absent or uncertain, honest AI attribution may be a safer choice.

\section{Data Availability Statement}
The data that support the findings of this study are available from the corresponding author, CM, upon reasonable request.

\section{Ethics Statement}
Study design was approved by MIT institutional IRB. 

\section{Generative AI Statement}
The author(s) declared that generative AI was used in the creation of this manuscript to suggest improvements for clarity. Generative AI was used in this study for participant feedback as a part of the study methodology. The author(s) verify that they are fully responsible for the content of this article. All intellectual and analytical contributions were made by the author(s).



\printbibliography

\end{document}